\documentclass[11pt]{scrartcl}  
\usepackage[left=1in, right=1in, top=1in, bottom=1in]{geometry}
\usepackage{mathptmx} 
\usepackage{multirow,graphicx,physics,hyperref,amsmath,amssymb,graphicx,mathalfa,mathtools,colortbl,hhline,bbold}
\numberwithin{equation}{section}
\hypersetup{colorlinks=true, citecolor=blue}
\usepackage{array}

\usepackage{longtable,caption}
\newcolumntype{L}[1]{>{\raggedright\let\newline\\\arraybackslash\hspace{0pt}}m{#1}}
\newcolumntype{C}[1]{>{\centering\let\newline\\\arraybackslash\hspace{0pt}}m{#1}}
\newcolumntype{R}[1]{>{\raggedleft\let\newline\\\arraybackslash\hspace{0pt}}m{#1}}
\usepackage[framemethod=TikZ]{mdframed}
\mdfdefinestyle{sid}{%
    linecolor=black,
    outerlinewidth=1.0pt,
    roundcorner=7pt,
    innerrightmargin=15pt,
    innerleftmargin=15pt,
    backgroundcolor=gray!30
		}	
\definecolor{Gray}{gray}{0.8}
\definecolor{MyBlue}{rgb}{0.0,0.0,0.9}
\definecolor{MyRed}{rgb}{0.0,0.9,0.0}
\colorlet{Bluee}{MyBlue!6}
\colorlet{Redd}{MyRed!1}
\DeclareMathAlphabet{\mathcal}{OMS}{cmsy}{m}{n} 

\usepackage{xcolor}
\colorlet{sectioncolor}{blue!20}
\colorlet{subsectioncolor}{orange!70}
\colorlet{subsubsectioncolor}{green!40}
\makeatletter
\renewcommand\sectionlinesformat[4]{%
  \colorbox{#1color}{%
    \parbox[t]{\dimexpr\textwidth-2\fboxsep\relax}{%
      \raggedsection\color{black}\@hangfrom{#3}{#4}%
}}}
\makeatother
\usepackage{blindtext}
\title{\textbf{Large-party limit of topological entanglement entropy in Chern-Simons theory}}
\date{}
\author{\textbf{\emph{Simran Sain}} and \textbf{\emph{Siddharth Dwivedi}}\thanks{Email: siddharth.dwivedi@curaj.ac.in, 2025phdphy002@curaj.ac.in }\\\\ Department of Physics, School of Physical Sciences, \\ Central University of Rajasthan, Ajmer, 305817, India}
\begin{document}
\maketitle
\begin{abstract}
We investigate the topological entanglement entropy of quantum states arising in the context of three-dimensional Chern-Simons theory with compact gauge group $G$ and Chern-Simons level $k$. We focus on the quantum states associated with the $T_{dm,dn}$ torus link complements, which is a $d$-party pure quantum state, and analyze its large-party limit, i.e., $d\to \infty$ limit. We show that the entanglement measures in this limit will receive contributions only from the Abelian anyons, and non-Abelian sectors are suppressed in the large-party limit. Consequently, the large-party limiting value of the entanglement entropy has an upper bound of $\ln |Z_G|$, where $|Z_G|$ is the order of the center of the group $G$. As an explicit example, we perform quantitative analysis for the simplest case of the SU(2) group and $T_{d,dn}$ torus link to obtain the large-party limit of the entanglement entropy. We further investigate the semiclassical ($k \to \infty$) limit of the entropies after taking the large-party limit for this particular example. 
\end{abstract}

\hypersetup{linkcolor=blue}
\tableofcontents
\newpage
\section{Introduction} \label{sec1}
The underlying theory for our study will be the (2+1)-dimensional Chern-Simons theory with a compact gauge group $G$ and the Chern-Simons level $k$, which is a non-negative integer. This theory is defined on a 3-manifold $M$ with action given by \cite{Witten:1988hf}
\begin{equation}
S(A) = \frac{k}{4\pi} \int_M \text{Tr}\left(A \wedge dA + \frac{2}{3} A \wedge A \wedge A \right) ~,
\label{CSaction}
\end{equation}
where the gauge field $A = A_{\mu}dx^{\mu}$ is a connection on the trivial $G$ bundle over $M$. This theory has gauge-invariant operators, which are called Wilson lines. For an oriented knot $\mathcal{K}$ embedded in $M$, the Wilson line is defined by taking the trace of the holonomy of $A$ around $\mathcal{K}$:
\begin{equation}
W_{R}(\mathcal{K}) = \text{Tr}_R \,P\exp\left(i\oint_{\mathcal{K}} A \right) ~,
\end{equation} 
where the trace is taken under the representation $R$ of SU($N$). One can compute the partition function by integrating over the gauge invariant classes of connections, which results in a topological invariant of $M$. When there are no knots or links, the partition function $Z(M)$ of the bare manifold is given as:
\begin{equation}
Z(M) = \int e^{i S(A)} dA ~.
\end{equation}
When there are knot insertions in $M$, one must insert the corresponding Wilson loop operators in the integral. Given a link $\mathcal{L}$ made of disjoint oriented knot components, i.e. $\mathcal{L} = \mathcal{K}_1 \sqcup \mathcal{K}_2 \sqcup \ldots \sqcup \mathcal{K}_d$, the partition function of $M$ in the presence of $\mathcal{L}$ can be obtained by modifying the path integral as:
\begin{equation}
Z(M; \mathcal{L}) = \int e^{i S(A)}\, W_{R_1}(\mathcal{K}_1)\ldots W_{R_d}(\mathcal{K}_d) \, dA ~.
\end{equation}  
In this article, we will consider manifolds $M$ with boundary $\Sigma$. For this, we must incorporate the boundary condition $A\rvert_{\Sigma} = Q$ while performing the path integral. In such cases, the partition function is interpreted as the wave function of a quantum state, which is given as:
\begin{equation}
\ket{\Psi} \equiv Z_Q(M; \mathcal{L}) = \int_{A\rvert_{\Sigma} = Q} e^{i S(A)}\, W_{R_1}(\mathcal{K}_1)\ldots W_{R_n}(\mathcal{K}_n) \, dA ~.
\end{equation}
This quantum state is an element of the Hilbert space $\mathcal{H}_{\Sigma}$ associated with the boundary $\Sigma$. In this work, we will be dealing with the torus boundary, i.e., $\Sigma=T^2$. A basis for the Hilbert space $\mathcal{H}_{T^2}$ is obtained by performing the Chern-Simons path integral on a solid torus with Wilson lines inserted along its non-contractible cycle. These Wilson lines can carry only certain allowed representations of the group $G$ for a given level $k$. These are called the `integrable representations'. For example, for SU($N$) group, the integrable representations are those representations $[a_1,a_2,\ldots,a_{N-1}]$ which satisfy $0 \leq a_1+a_2+\ldots+a_{N-1} \leq k$ and $a_i \geq 0$. Here, $a_i$ are the Dynkin labels of the highest weight of the representation.\footnote{For the readers who are familiar with the Young tableau notation, the representation $R=[a_1,a_2,\ldots,a_{N-1}]$ of SU($N$) will correspond to a Young tableau whose $i^{\text{th}}$ row has $\ell_i = a_i+a_{i+1}+\ldots + a_{N-1}$ boxes.} Since there is a finite number of integrable representations at a given level, the Hilbert space $\mathcal{H}_{T^2}$ is finite-dimensional with a basis given as (for SU($N$) case):
\begin{equation}
\text{basis}(\mathcal{H}_{T^2}) = \{\ket{e_R} : R=[a_1,a_2,\ldots,a_{N-1}] \,\, ;\,\,   a_1+a_2+\ldots+a_{N-1} \leq  k \,\,\& \,\,  a_i \geq 0 \}~,
\label{basisHT2}
\end{equation}
where the state $\ket{e_R}$ is obtained by performing the path integral on a solid torus with Wilson line carrying representation $R$ inserted along the non-contractible cycle of the solid torus.

When a manifold $M$ has multiple (say $d$) disjoint torus boundaries, a generic quantum state will be a $d$-party state which lives in the tensor product of $d$ copies of $\mathcal{H}_{T^2}$ and can be written as: 
\begin{equation}
\ket{\Psi} = \sum_{R_1} \sum_{R_2} \cdots \sum_{R_d} C_{R_1,R_2,\ldots,R_d} \, \ket{e_{R_1},e_{R_2},\ldots,e_{R_d}} ~,
\label{Genstate}
\end{equation}
where $\ket{e_{R_1},e_{R_2},\ldots,e_{R_d}}$ is the shorthand way of writing $\ket{e_{R_1}} \otimes \ket{e_{R_2}} \otimes \ldots \otimes \ket{e_{R_d}}$ and $C_{R_1,R_2,\ldots,R_d}$ are complex numbers which carry the topological information of $M$. The sum $R_i$ is the sum over all the integrable representations of group $G$ with level $k$ in the context of the $i^{\text{th}}$ torus boundary. Such multi-boundary quantum states living in the tensor products of $\mathcal{H}_{T^2}$ have been extensively studied, and we refer the readers to \cite{Balasubramanian:2016sro,Dwivedi:2017rnj,Balasubramanian:2018por,Salton:2016qpp,Hung:2018rhg,Zhou:2019ezk,Dwivedi:2020jyx,Dwivedi:2020rlo,Dwivedi:2021dix,Dwivedi:2024gzg,Saini2025,Cummings:2025zfe,Yuan:2025dgx,Munizzi:2025suf,Akella:2026xza} for advances in this field of study. These references examine various aspects of topological entanglement measures for such multi-boundary states.

In this work, we shall investigate a new aspect, namely, the large-party limit of the topological entanglement entropy. Our focus will be on the quantum state associated with the link complement $M =S^3 \backslash T_{p,q}$ where $T_{p,q}$ is a generic torus link. Assuming that $d=\text{gcd}(p,q)$, the link complement $S^3 \backslash T_{p,q}$ will have $d$ disjoint torus boundaries. So the quantum state associated with $S^3 \backslash T_{p,q}$ will naturally be a $d$-party state, which we shall denote as $\ket{T_{p,q}}$. Our goal will be to analyze the large-party, i.e, $d \to \infty$ limit of the entanglement structure of this state for a generic group $G$ and level $k$. We will provide a detailed quantitative analysis and results of one example of $\ket{T_{d,dn}}$ for the SU(2) group. For this example, we will also obtain the semiclassical limit (i.e. $k \to \infty$) after we have obtained the large-party (i.e. $d \to \infty$) limit. Our results and findings are presented in Section \ref{sec2}, and we conclude in Section \ref{sec3} with a discussion on future directions.
\section{Generic $d$-party torus link state $\ket{T_{dm,\,dn}}$} \label{sec2}
Let us consider the torus link state $\ket{T_{p,q}}$ where $d=\text{gcd}(p,q)$. Without loss of generality, set $p=dm$ and $q=dn$ with $\text{gcd}(m,n)=1$. The torus link $T_{dm,dn}$ is made of $d$ torus knots $T_{m,n}$ such that the linking number between any pair of knots is $mn$. This is a generic $d$-party torus link state which lives in the tensor product of $d$ copies of $\mathcal{H}_{T^2}$ and can be linearly expanded as:
\begin{equation}
\ket{T_{dm,\,dn}} = \sum_{R_1} \sum_{R_2} \cdots \sum_{R_d} C_{R_1,R_2,\ldots,R_d} \, \ket{e_{R_1},e_{R_2},\ldots,e_{R_d}} ~,
\label{Linkstate}
\end{equation}
There is a nice topological way of extracting the values of the coefficients $C_{R_1,R_2,\ldots,R_d}$ as explained in \cite{Balasubramanian:2016sro} and illustrated in \cite{Dwivedi:2020rlo,Dwivedi:2024gzg}, and in this case, they will be given as:
\begin{equation}
C_{R_1,R_2,\ldots,R_d} = Z(S^3;[T_{dm,\,dn}]_{R_1,R_2,\ldots,R_d}) ~,
\end{equation}
where the right hand side is the $S^3$ partition function in the presence of the torus link $T_{dm,\,dn}$ where the $d$ knots carry the representations $R_1,R_2,\ldots R_d$ of the gauge group $G$. These partition functions have already been obtained in the works of \cite{Stevan:2010jh,Brini:2011wi} and can be written in terms of the modular $\mathcal{S}$ and $\mathcal{T}$ matrices, which form the unitary representation of the modular group SL(2, $\mathbb{Z}$).
The partition function is given as:
\begin{equation}
 Z(S^3;[T_{dm,\,dn}]_{R_1,R_2,\ldots,R_d}) = \sum_{P,Q,R}\frac{(\mathcal{S}_{PQ})^*\,\mathcal{S}_{0R} \, (\mathcal{T}_{RR})^{n/m}\, \mathcal{X}_{QR}(m)}{(\mathcal{S}_{0P})^{d-1}} \left(\mathcal{S}_{R_1P} \, \mathcal{S}_{R_2P} \ldots \mathcal{S}_{R_dP} \right) ~.
 \label{TorusLinkPF}
\end{equation}
The coefficients $\mathcal{X}_{QR}(m)$ are integers that occur in the expansion of the traces of powers of the holonomy operator $U$ as:
\begin{equation}
\text{Tr}_{Q}(U^m) = \sum_{R} \mathcal{X}_{QR}(m)\, \text{Tr}_{R}(U) ~.
\label{AdamsGen}
\end{equation}
These coefficients can be obtained by performing the Adams operation on the characters associated with the irreducible representations of the group $G$. For the SU(2) group, these coefficients have been explicitly computed in \cite{Dwivedi:2020rlo}. A computationally friendly formula for the matrix elements of $\mathcal{S}$ and $\mathcal{T}$ can be found in \cite{Dwivedi:2017rnj}. Although the quantum state \eqref{Linkstate} after substituting \eqref{TorusLinkPF} looks cumbersome, it can be written in a cleaner fashion by changing the basis of individual Hilbert spaces. Since we do not want the entanglement measures to change, we must use unitary transformed basis states for each Hilbert space. Since $\mathcal{S}$ matrix is unitary, we can invoke the following basis change for the $i^{\text{th}}$ Hilbert space:
\begin{equation}
\ket{e_{R_i}} = \sum_{P_i} (\mathcal{S}_{R_i P_i})^* \ket{f_{P_i}} ~.
\end{equation}
It is a simple exercise to see that in the new basis $\ket{f_{P_i}}$ of the $i^{\text{th}}$ Hilbert space, the quantum state can be linearly expanded as:
\begin{equation}
\ket{T_{dm,\,dn}} = \sum_{P,Q,R}  \frac{(\mathcal{S}_{PQ})^*\,\mathcal{S}_{0R} \, (\mathcal{T}_{RR})^{n/m}\, \mathcal{X}_{QR}(m)}{(\mathcal{S}_{0P})^{d-1}} \, \underbrace{\ket{f_{P},\,f_{P},\,\ldots,\,f_{P}}}_{d} ~.
\label{LinkstateNewBasis}
\end{equation}
It can be written in a compact way as:\footnote{The states that can be written in this fashion are called the GHZ-like states, where a partial trace on any subsystem results in a separable reduced system. For a detailed discussion on what link complements result in a GHZ-like state, see \cite{Balasubramanian_2025}.}
\begin{equation}
\ket{T_{dm,\,dn}} = \sum_{P}  \frac{(\mathcal{S}^*\, \mathcal{X}(m)\,\mathcal{T}^{n/m}\, \mathcal{S})_{P0}}{(\mathcal{S}_{0P})^{d-1}} \, \ket{f_{P},\,f_{P},\,\ldots,\,f_{P}} ~.
\label{LinkstateCompact}
\end{equation}
In the next section, we will analyze the entanglement structure of this state in the $d \to \infty$ limit.
\subsection{Entanglement measures in large-party limit for generic group} \label{sec21}
Let us study the quantum state in \eqref{LinkstateCompact} for a compact gauge group $G$ and level $k$. Our first step must be to normalize the state. But before that, we would like to convert the isolated factor of $\mathcal{S}_{0P}$ into a quantum dimension using the formula $\text{dim}_qP = \mathcal{S}_{0P}/\mathcal{S}_{00}$. We will see the implications of this conversion later. So after normalization, the state \eqref{LinkstateCompact} will become:
\begin{equation}
\ket{T_{dm,\,dn}} = \frac{1}{\sqrt{\text{NF}}}\, \sum_{P}  \frac{(\mathcal{S}^*\, \mathcal{X}(m)\,\mathcal{T}^{n/m}\, \mathcal{S})_{P0}}{(\text{dim}_qP)^{d-1}} \, \ket{f_{P},\,f_{P},\,\ldots,\,f_{P}} ~,
\label{LinkstateCompactNormalized}
\end{equation}
where NF is the normalization factor given as:
\begin{equation}
\text{NF} = \langle T_{dm,\,dn}| T_{dm,\,dn} \rangle = \sum_{R} \abs{\frac{(\mathcal{S}^*\, \mathcal{X}(m)\,\mathcal{T}^{n/m}\, \mathcal{S})_{R0}}{(\text{dim}_qR)^{d-1}}}^2 ~.
\end{equation}
The structure of the state immediately tells us that the total density matrix $\rho_{\text{total}}=\ket{T_{dm,\,dn}}\bra{T_{dm,\,dn}}$ will be a diagonal matrix of order $(n_0)^d$. Here $n_0 \equiv \text{dim } \mathcal{H}_{T^2}$ is the dimension of the Hilbert space for a given value of $k$, and it essentially counts the number of integrable representations of the group $G$ at level $k$. This counting can be explicitly found in \cite{Dwivedi:2017rnj}. Of all the diagonal elements, only the $n_0$ number of diagonal elements of $\rho_{\text{total}}$ will be non-zero, and others will be 0. In order to study the entanglement measures, we would like to bi-partition the total system as $(A|B) = (d'|d-d')$, where Alice's subsystem has $d'$ parties and Bob's subsystem has $(d-d')$ parties. The corresponding bi-partitioning of the total Hilbert space will be:
\begin{equation}
\underbrace{\mathcal{H}_{T^2} \otimes \mathcal{H}_{T^2} \otimes \mathcal{H}_{T^2}}_{d'} \,\otimes \, \underbrace{\mathcal{H}_{T^2} \otimes \mathcal{H}_{T^2} \otimes \mathcal{H}_{T^2}}_{d-d'}
\end{equation}
Tracing out Bob's system will give us the reduced density matrix on Alice's system. The diagonal structure of the total density matrix ensures that the reduced density matrix $\rho_A$ will also be a diagonal matrix, with only $n_0$ non-zero entries and all other entries being zero. Since the diagonal entries are also the eigenvalues of $\rho_A$ and the zero eigenvalues do not contribute to the entanglement measures, the entanglement structure is insensitive to how many parties are traced out. This is a generic feature of torus link states and is a consequence of the GHZ-like entanglement structure of torus link states as discussed in \cite{Balasubramanian:2018por} and \cite{Balasubramanian_2025}. So without loss of generality, we can assume the bi-partition $(A|B) = (1|d-1)$. Now, tracing out $B$ will give a reduced density matrix of order $n_0$ whose eigenvalues will be given as:
\begin{equation}
\lambda_{R} = \frac{F_R}{\text{NF}} \quad;\quad F_R = \abs{\frac{(\mathcal{S}^*\, \mathcal{X}(m)\,\mathcal{T}^{n/m}\, \mathcal{S})_{R0}}{(\text{dim}_qR)^{d-1}}}^2 ~,
\label{EigenvaluesRDM}
\end{equation} 
where each integrable representation gives one eigenvalue. Our goal is to study the large-party limit, i.e., $d \to \infty$ limit of the entanglement entropy. Let us first examine how the eigenvalues themselves behave in this limit. We will assume that $m$ and $n$ are finite integers and only $d \to \infty$. From \eqref{EigenvaluesRDM}, we see that the explicit dependence of $d$ appears only in the term $(\text{dim}_qR)^{d-1}$. Let us rewrite \eqref{EigenvaluesRDM} by separating out the $d$ dependence:
\begin{equation}
\lambda_{R} = \frac{M_R}{\sum\limits_P M_P\, (\text{dim}_qR/\text{dim}_qP)^{2d-2}}  \quad;\quad M_R = \abs{(\mathcal{S}^*\, \mathcal{X}(m)\,\mathcal{T}^{n/m}\, \mathcal{S})_{R0}}^2 ~.
\end{equation}
Now, in our theory, we will have both Abelian and non-Abelian anyons at a given level $k$. The Abelian anyons correspond to those integrable representations $R=R_a$ which have unit quantum dimension, i.e. $\text{dim}_qR_a = 1$. For non-Abelian anyons at a given value of $k$, the quantum dimension is always greater than 1. So, we can write the following:
\begin{equation}
\sum_P M_P\, \left(\frac{\text{dim}_qR}{\text{dim}_qP}\right)^{2d-2} = \sum_{P\, \in \,\text{Abelian}} M_P\, \left(\text{dim}_qR\right)^{2d-2} + \sum_{P\, \in \, \text{non-Abelian}} M_P\, \left(\frac{\text{dim}_qR}{\text{dim}_qP}\right)^{2d-2} ~.
\end{equation}
Therefore, for large $d$, we will get the following limit: 
\begin{equation}
\lim_{d \to \infty}\,\sum_P M_P\, \left(\frac{\text{dim}_qR}{\text{dim}_qP}\right)^{2d-2} = \begin{cases}
    \sum\limits_{P\, \in \,\text{Abelian}} M_P  & \text{, \quad when $R \in $ Abelian} \\[0.3cm]  \infty & \text{, \quad when $R \in $ non-Abelian}
\end{cases} ~.
\end{equation}
Thus, the large-party limit of the eigenvalues will be given as:
\begin{equation}
\lambda_{R}^{\text{LP}} \,\,\equiv \,\, \lim_{d \to \infty}\lambda_{R} \,\,=  \,\, \begin{cases}
    \dfrac{M_R}{\sum_{P\, \in \,\text{Abelian}}M_P}  & \text{, \quad when $R \in $ Abelian} \\[0.3cm]  0 & \text{, \quad when $R \in $ non-Abelian}
\end{cases} ~,
\end{equation}
where it is assumed that $M_R$ itself does not diverge for any $R$ and any $k$, which is true and can be easily verified using numerical checks. The superscript `LP' in $\lambda_{R}^{\text{LP}}$ is a notation to say that these are the eigenvalues in the large-party limit. Thus, we see that only those eigenvalues that correspond to Abelian anyons survive. So we get our result:\\ \\
\textbf{Result-1:} \emph{The topological entanglement measures for the state $\ket{T_{dm,\,dn}}$ in the large-party limit, i.e., $d \to \infty$ limit, receive contributions only from the Abelian anyons of the theory, and the non-Abelian sectors are suppressed in the large-party limit.}\\ \\
We also know that the Abelian anyons are in one-to-one correspondence with the center $Z_G$ of the gauge group $G$. Hence, the number of Abelian anyons is equal to the order of $Z_G$. This gives us the following result.\\ \\
\textbf{Result-2:} \emph{The large-party limit, i.e. $d \to \infty$ limit, of the entanglement entropy for the state $\ket{T_{dm,\,dn}}$ will satisfy the bound:} $0 \leq \text{EE}_{\text{LP}} \leq \ln |Z_G|$.\\ \\
For example, if we consider the SU($N$) group with level $k$, the Abelian anyons correspond to the following representations:
\begin{equation}
\text{Abelian anyons} \in \left\{ [0,0,\ldots,0] \,\,,\,\, [k,0,0,\ldots,0] \,\,,\,\, [0,k,0,\ldots,0] \,\,,\,\, \ldots \,\,,\,\, [0,0,\ldots,0,k] \right\}  ~.
\end{equation}
Hence, in the large-party limit, only those eigenvalues $\lambda_R$ will survive for which $R$ is one of the representations listed above. The value of the entropy $\text{EE}_{\text{LP}}$ will, of course depend on $k,$ but it will always satisfy the bound $0 \leq \text{EE}_{\text{LP}} \leq \ln N$ for any $k$.

In the following, we will perform a quantitative analysis for one example: the link state is $\ket{T_{d,\,dn}}$ and the gauge group is the SU(2) group. For this particular example, we will study the large-party entanglement entropy for finite $k$ and will also analyze its large $k$ behavior.
\subsection{An example: Quantum state $\ket{T_{d,\,dn}}$ and SU(2) group}
Let us consider the $m=1$ case, so that $n$ can be taken as any positive integer. From the equation \eqref{AdamsGen}, we can see that $\mathcal{X}_{QR}(1) = \delta_{QR}$. Hence,  $\mathcal{X}(1)$ will be an identity matrix. Therefore, we can write:
\begin{equation}
M_R = \abs{(\mathcal{S}^*\,\mathcal{T}^{n}\, \mathcal{S})_{R0}}^2 ~.
\end{equation}
The Abelian anyons for the SU(2) group and level $k$ are $R=[0]$ and $R=[k]$, which are the trivial and spin $k/2$ representations, respectively. Hence, in the $d \to \infty$ limit, only two eigenvalues survive which are:
\begin{equation}
\lambda_{0} = \frac{M_{0}}{M_{0}+M_{k}} \quad;\quad \lambda_{k} = \frac{M_{k}}{M_{0}+M_{k}} ~.
\end{equation}
The entanglement entropy will be given as:
\begin{equation}
\text{EE}_{\text{LP}} = -(\lambda_{0} \ln \lambda_{0} + \lambda_{k} \ln \lambda_{k}) ~.
\end{equation}
In the remainder of this section, we will analyze this entanglement entropy for finite $k$ as well as for large $k$.
\subsubsection{Large-party entanglement entropy for finite $k$}
\begin{figure}[htb]
    \centering
    \includegraphics[width=1.0\textwidth]{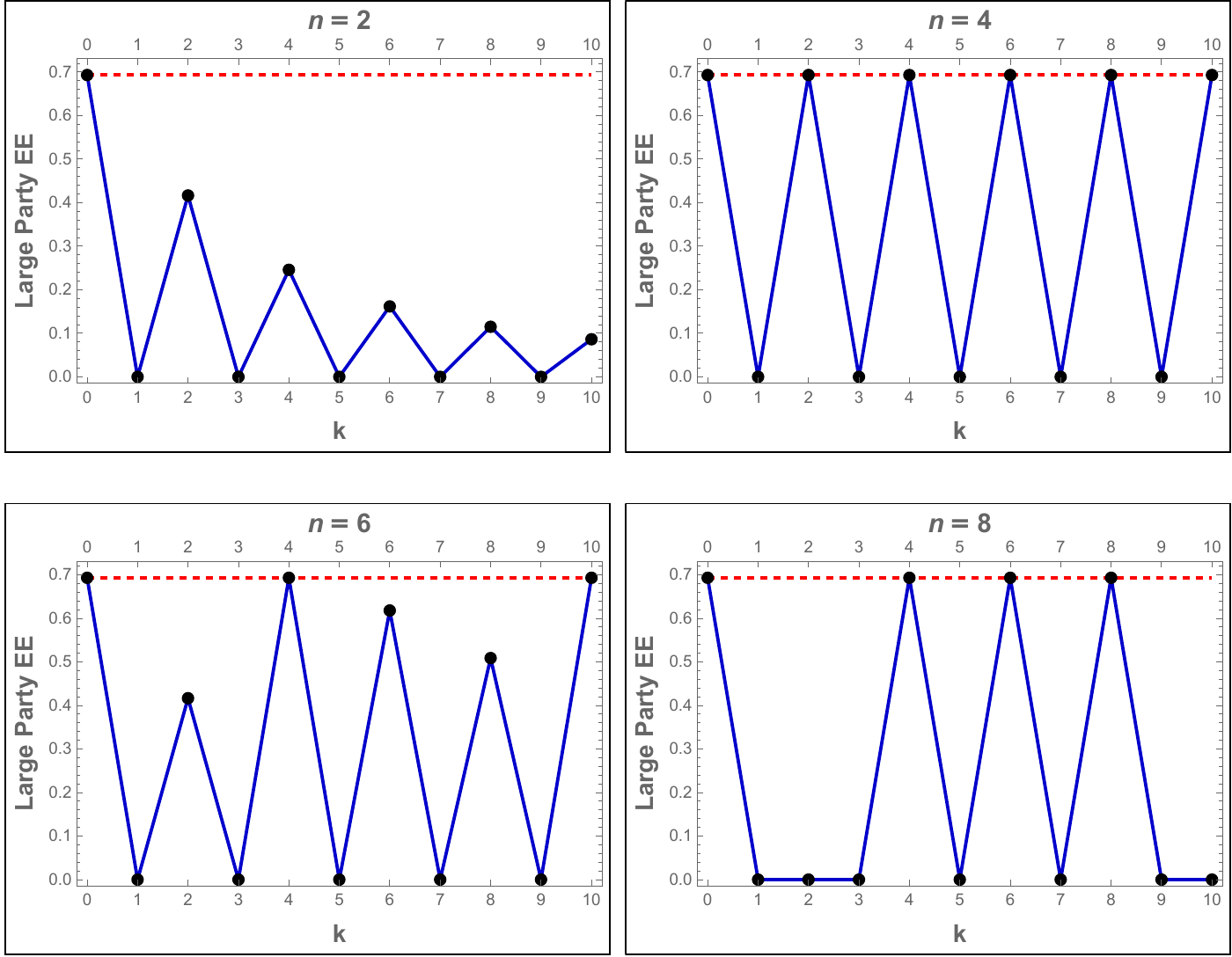}
    \caption{Variation of $\text{EE}_{\text{LP}}$ vs $k$ for quantum states $\ket{T_{d,\,dn}}$ and SU(2) group. The plots are shown for $n=2,4,6,8$. The dotted red line indicates the maximum possible value, which is $\ln 2$.}
    \label{EEPLPlot}
\end{figure}

The modular $\mathcal{S}$ and $\mathcal{T}$ matrix elements for SU(2) group are given below:
\begin{equation}
\mathcal{S}_{ab} = \sqrt{\frac{2}{k+2}}\, \sin\left[\frac{(a+1)(b+1)\pi}{k+2}\right] \quad;\quad \mathcal{T}_{ab} = e^{\frac{i \pi a (a+2)}{2 (k+2)}} e^{-\frac{i \pi  k}{4 (k+2)}}\, \delta_{ab} ~.
\end{equation}
Using this, we can numerically compute the numbers $M_0$ and $M_k$ and obtain the eigenvalues $\lambda_0$ and $\lambda_k$ and hence the large-party entanglement entropy $\text{EE}_{\text{LP}}$. Although we do not have a closed-form expression for the entropy as a function of $k$, we present certain observations in the following.\\ \\
\textbf{Result-3:} \emph{For the state $\ket{T_{d,\,dn}}$ and the group SU(2), the large-party limit of the entanglement entropy shows the following behavior:\\
$\bullet$ The states $\ket{T_{d,\,dn}}$ are maximally entangled large-party states for odd $n$. \\
$\bullet$ The states $\ket{T_{d,\,dn}}$ are separable large-party states for even $n$ and odd $k$.
} \\ \\
The first part of the result is a consequence of the fact that for odd values of $n$, we will have $M_0=M_k$ for any value of $k$, which we have checked numerically. Therefore, $\lambda_0=\lambda_k=1/2$ giving the maximum entropy $\text{EE}_{\text{LP}}=\ln 2$. The second part of the result is because for even $n$ and odd $k$ values, one of the $M_0$ or $M_k$ vanishes. As a result, one of the eigenvalues is 0, and the other is 1, resulting in vanishing entropy.
For even $n$ and even $k$ values, the entropy varies between 0 and $\ln 2$ with no fixed pattern at first glance. We present the plots of the $\text{EE}_{\text{LP}}$ vs $k$ for $n=2,4,6,8$ in the Figure \ref{EEPLPlot}.
\subsubsection{Large-party entanglement entropy at large $k$}
Now, let us analyze the semiclassical, i.e., large $k$ behavior of the large-party entanglement entropy. For this, we must first obtain the large $k$ asymptotics of $M_0$ and $M_k$. Let us define $M_0 = |L_0|^2$ and $M_k = |L_k|^2$ so that we will have:
\begin{equation}
 L_{0} = (\mathcal{S}^*\,\mathcal{T}^{n}\, \mathcal{S})_{00} \quad;\quad L_{k} = (\mathcal{S}^*\,\mathcal{T}^{n}\, \mathcal{S})_{k0} ~.
\end{equation}
We can rewrite these functions as follows:
\begin{equation}
L_0 = \sum_{a=0}^{k} \mathcal{S}_{0 a} T_{aa}^n \mathcal{S}_{a0}  = \sum_{a=0}^{k} \mathcal{S}_{0a}^2 T_{aa}^n \quad;\quad  L_k = \sum_{a=0}^{k} \mathcal{S}_{ka} T_{aa}^n \mathcal{S}_{a0}  = \sum_{a=0}^{k} (-1)^a\,\mathcal{S}_{0a}^2 T_{aa}^n ~,
\end{equation}
where we have used the fact $\mathcal{S}_{ka}=(-1)^a\mathcal{S}_{0a}$. Now we will substitute the $\mathcal{S}$ and $\mathcal{T}$ matrix elements of SU(2) group  and define $K=(k+2)$ so that we can write:
\begin{align}
L_0 &= \frac{2\,e^{-\frac{i \pi n (K-2)}{4K}}}{K}\sum_{a=0}^{K-2} \sin^2\left[\frac{(a+1)\pi}{K}\right]\, e^{\frac{i \pi n a (a+2)}{2K}} \nonumber \\ 
L_k &= \frac{2\,e^{-\frac{i \pi n (K-2)}{4K}}}{K}\sum_{a=0}^{K-2} \sin^2\left[\frac{(a+1)\pi}{K}\right]\,e^{i\pi a} e^{\frac{i \pi n a (a+2)}{2K}} ~.
\end{align}
In the large $K$ limit, we can convert the sum in $L_0$ and $L_k$ to a sum of integrals using the Poisson summation formula, which we have discussed in Appendix A. Defining $x\equiv (a+1)/K$ and using the formula \eqref{LargeKformula}, we can write:
\begin{align}
L_0 &\,\, \sim \,\, 2\,e^{-\frac{i \pi n (K-2)}{4K}}\,e^{-\frac{i \pi  n}{2K}} \sum_{s=-\infty}^{\infty}\, \int_{0}^1 \sin^2(\pi  x)\, e^{iK \left(\frac{n\pi x^2}{2}-2\pi s x\right)} dx \nonumber \\ 
L_k &\,\, \sim \,\, 2\,e^{-\frac{i \pi n (K-2)}{4K}}e^{i \pi }\,e^{-\frac{i \pi  n}{2K}} \sum_{s=-\infty}^{\infty}\, \int_{0}^1 \sin^2(\pi  x)\, e^{iK \left(\frac{n\pi x^2}{2}+\pi x-2\pi s x\right)} dx ~.
\end{align}
So now, our aim is to obtain the large $K$ asymptotics of the integrals 
\begin{align}
I_s &= \int_{0}^1 \sin^2 (\pi  x)\, e^{iK\, \phi_s(x)}\, dx \quad;\quad \phi_s(x) = \frac{n\pi x^2}{2}-2\pi s x \nonumber \\
J_s &= \int_{0}^1 \sin^2 (\pi  y)\, e^{iK\, \Phi_s(y)}\, dy \quad;\quad \Phi_s(y) = \frac{n\pi y^2}{2}+\pi y-2\pi s y~.
\end{align}
We will use the method of the stationary phase for this, which we have discussed in Appendix B. To apply the method of the stationary phase, we need to find the stationary points in the two cases. For the integrals $I_s$ and $J_s$, the stationary points are $x_0$ and $y_0$ at which $\phi_s'(x_0)=0$ and $\Phi_s'(y_0)=0$ respectively. These points are given as:
\begin{equation}
x_0 = \frac{2s}{n} \quad;\quad  y_0 = \frac{2s-1}{n} ~.
\end{equation}
Let us consider the integral $I_s$ first. Since $n$ is a fixed positive integer, the stationary points that lie in the integration range $[0,1]$ will occur only for $s=0,1,2,\ldots,\left\lfloor \frac{n}{2}\right\rfloor$. Here $\left\lfloor a\right\rfloor$ is the Floor function, which gives the greatest integer less than or equal to $a$. Since the leading order contribution in the large $K$ asymptotics of $I_s$ comes only from the stationary points that are within the integration range $[0,1]$, we will only consider the contribution of the integrals $I_s$ where $s=0,1,2,\ldots,\left\lfloor \frac{n}{2}\right\rfloor$. Thus our original function $L_0$ and hence $M_0$ can be approximated as:   
\begin{equation}
L_{0} \,\, \sim \,\, 2\,e^{-\frac{i \pi n (K-2)}{4K}}\,e^{-\frac{i \pi  n}{2K}} \sum_{s=0}^{\left\lfloor \frac{n}{2}\right\rfloor} I_s \quad \Longrightarrow\quad M_{0} \,\, \sim \,\, 4\abs{\sum_{s=0}^{\left\lfloor n/2\right\rfloor} I_s}^2 ~.
 \end{equation}
Now consider the integral $J_s$. Here the stationary points that lie in the integration range $[0,1]$ will occur only for $s=1,2,\ldots,\left\lfloor \frac{n+1}{2}\right\rfloor$. Hence, we will have:
\begin{equation}
L_{k} \,\, \sim \,\, 2\,e^{-\frac{i \pi n (K-2)}{4K}}\,e^{i\pi}\,e^{-\frac{i \pi  n}{2K}} \sum_{s=1}^{\left\lfloor \frac{n+1}{2}\right\rfloor} J_s \quad \Longrightarrow\quad M_{k} \,\, \sim \,\, 4\abs{\sum_{s=1}^{\left\lfloor \frac{n+1}{2}\right\rfloor} J_s}^2 ~.
\end{equation}
For each stationary point $x_0$ and $y_0$, we can apply equations \eqref{x0LeftSP}, \eqref{x0SP}, or \eqref{x0RightSP}, depending upon whether the stationary point coincides with the endpoints of the integration (which are 0 and 1), or if it is in the interval $(0,1)$. It should be noted here that the function $f(x)=\sin^2(\pi x)$ and its derivative $f'(x)$ vanish at both the endpoints, i.e., at $x=0$ as well as $x=1$. As a result, the boundary stationary points will only contribute at the order $k^{-3/2}$ in both $I_s$ and $J_s$. Also, since both $f(x)$ and $f'(x)$ have non-zero values for $x\in (0,1)$, the stationary points in the interval $(0,1)$ will contribute at the order $k^{-1/2}$ in both $I_s$ and $J_s$. 

There is another subtle point that needs to be mentioned here. In the equations \eqref{x0LeftSP}, \eqref{x0SP} and \eqref{x0RightSP}, we notice an exponential term $\exp[iK\phi(x_0)]$. In our case, this term will be $\exp[-2 i \pi (k+2) s^2/n]$ in $I_s$ and $\exp[-i \pi (k+2) (2s-1)^2/(2n)]$ in $J_s$ respectively. These are not overall phase factors, as they are $s$-dependent and hence the physical results will be sensitive to the value of $k$ modulo $n$. Therefore, we must consider $k \equiv r \, (\text{mod $n$})$ where $r=0,1,2,\ldots,(n-1)$. For each value of $r$, the large $k$ analysis may yield different leading-order coefficients (for the same $n$), and we must keep track of these distinct leading-order coefficients. Our analysis gives the following form of the large $k$ behavior of $M_0$ and $M_k$:
\begin{equation}
M_0 \sim  \begin{cases}
    \dfrac{C(r,n)}{k} & \text{for $n\geq 3$} \\[0.3cm] \dfrac{2\pi ^2}{k^3} & \text{for $n=1$} \\[0.3cm] \dfrac{\pi ^2 \left((-1)^r+1\right)^2}{4 k^3} & \text{for $n=2$}
\end{cases} \quad;\quad M_k \sim  \begin{cases}
    \dfrac{D(r,n)}{k} & \text{for $n\geq 3$} \\[0.3cm] \dfrac{2\pi ^2}{k^3} & \text{for $n=1$} \\[0.3cm] \dfrac{4}{k} & \text{for $n=2$}
\end{cases} ~.
\label{M0Mkasymp}
\end{equation}
Here is a brief explanation of these results. For $n=1$, we have $M_0 \sim 4\abs{I_0}^2$ and $M_k \sim 4\abs{J_1}^2$. For this case, the stationary point of $I_0$ is $x_0=0$, and the stationary point of $J_1$ is $y_0=1$. Since these are the boundary points and $\sin^2(\pi x)$ and its derivative vanish at these points, the integrals $I_0$ and $J_1$ both go as $k^{-3/2}$, giving a $k^{-3}$ behavior of $M_0$ and $M_k$ for $n=1$. The relevant formulae here will be \eqref{x0LeftSP} for $I_0$ and \eqref{x0RightSP} for $J_1$.

For $n=2$, we will have: $M_0 = 4\abs{I_0+I_1}^2$ and $M_k = 4\abs{J_1}^2$. For the case of $M_0$, the stationary points of $I_0$ and $I_1$ are $x_0=0$ and $x_0=1$ respectively. Both being the boundary points, we will get the expected $k^{-3}$ behavior of $M_0$. However, as discussed earlier, the asymptotics will be sensitive to $k \equiv r \, (\text{mod 2})$. So we may get different results for $r=0$ (i.e., when $k$ is even) and for $r=1$ (i.e., when $k$ is odd). In fact, we see that for $r=1$, the leading order coefficient vanishes. This is a consequence of our earlier discussion (after Result-3) that for even $n$ and odd $k$, one of $M_0$ or $M_k$ identically vanishes, and for $n=2$, it is $M_0$ that vanishes for odd $k$. Coming back to $M_k = 4\abs{J_1}^2$ for $n=2$, the stationary point is $y_0 = 1/2$ which lies in $(0,1)$. This will give the $k^{-1}$ behavior of $M_k$. Note that in this case $D(0,2)=D(1,2)=4$ and hence we get the same leading order coefficient for both even and odd $k$ asymptotics.  

For $n \geq 3$, we will get the stationary points in the interval $(0,1)$ for both $I_s$ and $J_s$, in addition to the boundary stationary points, if any. For $n \geq 3$, the leading order asymptotics will be dominated only by the stationary points that lie in the interval $(0,1)$. This will give the $k^{-1}$ behavior of $M_0$ and $M_k$ for $n \geq 3$. We can also write the generic formula for the leading order coefficients $C(r,n)$ and $D(r,n)$. For this, we note that the stationary points that lie inside the interval $(0,1)$ are as follows:
\begin{alignat}{2}
x_0  &= \frac{2s}{n} & \quad,\quad & \text{with}\,\,  s=1,2,3,\ldots,\left\lfloor \frac{n-1}{2}\right\rfloor \nonumber \\
y_0  &= \frac{2s-1}{n}  & \quad,\quad & \text{with}\,\,  s=1,2,3,\ldots,\left\lfloor \frac{n}{2}\right\rfloor  ~.
\end{alignat}
Obtaining the asymptotics of $I_s$ and $J_s$ using the formula \eqref{x0SP} and summing over all the contributions, we obtain the leading order coefficients for $n\geq 3$ as:
\begin{equation}
C(r,n) =  \dfrac{8}{n} \abs{\sum\limits_{s=1}^{\left\lfloor \frac{n-1}{2}\right\rfloor}  e^{-\frac{2 i \pi  (r+2) s^2}{n}} \sin ^2\left(\frac{2 \pi  s}{n}\right) }^2 \,\,;\,\, D(r,n) =  \dfrac{8}{n} \abs{\sum\limits_{s=1}^{\left\lfloor \frac{n}{2}\right\rfloor}  e^{-\frac{i \pi  (r+2) (1-2 s)^2}{2 n}} \sin ^2\left(\frac{\pi -2 \pi  s}{n}\right)}^2 ~.
\label{CandD}
\end{equation} 
Now that we know the large $k$ asymptotics of $M_0$ and $M_k$, we can obtain the large $k$ limits of the eigenvalues: 
\begin{equation}
\lambda_{0}^{\infty,\,\text{LP}} \equiv  \lim_{k \to \infty}\lambda_{0}^{\text{LP}} = \lim_{k \to \infty} \left(\frac{M_{0}}{M_{0}+M_{k}}\right) \quad;\quad \lambda_{k}^{\infty,\,\text{LP}} \equiv \lim_{k \to \infty}\lambda_{k}^{\text{LP}} = \lim_{k \to \infty} \left(\frac{M_k} {M_{0}+M_{k}}\right)  ~.
\end{equation}
From these eigenvalues, we can also calculate the large $k$ limit of the entanglement entropy as:
\begin{equation}
\text{EE}_{\text{LP}}^{\infty} \equiv \lim_{k \to \infty} \text{EE}_{\text{LP}} = - \left[\lambda_{0}^{\infty,\,\text{LP}} \ln\left( \lambda_{0}^{\infty,\,\text{LP}} \right) + \lambda_{k}^{\infty,\,\text{LP}} \ln\left( \lambda_{k}^{\infty,\,\text{LP}} \right) \right] ~.
\end{equation}
Note that if $M_0$ and $M_k$ have different $k$ asymptotics or if either is 0, then one of the eigenvalues will vanish, resulting in a vanishing limit of the entropy. As an example, for $n=2$ and $r=0$, we have $M_0 \sim \pi^2/k^3$ and $M_k \sim 4/k$. For $n=2$ and $r=1$, we will have $M_0 = 0$ and $M_k \sim  4/k$. Thus for $n=2$, we will have $\text{EE}_{\text{LP}}^{\infty} = 0$. For $n=1$, both $M_0 \sim 2\pi^2/k^3$ and $M_k \sim 2\pi^2/k^3$, giving the maximum entropy $\text{EE}_{\text{LP}}^{\infty} = \ln2$. For $n\geq 3$, we have $M_0 \sim C(r,n)/k$ and $M_k \sim D(r,n)/k$, where the leading order coefficients $C(r,n)$ and $D(r,n)$ are given in \eqref{CandD}. Finally, we are ready to present our results.   \\ \\
\textbf{Result-4:} \emph{The semiclassical limit of the large-party entanglement entropy of the state $\ket{T_{d,\,dn}}$ for the SU(2) group is given below:}
\begin{equation}
\text{EE}_{\text{LP}}^{\infty} \equiv \lim_{k \to \infty} \text{EE}_{\text{LP}} = \ln\left[C(r,n) + D(r,n) \right] - \frac{C(r,n) \ln C(r,n) + D(r,n) \ln D(r,n)}{C(r,n) + D(r,n)} ~.
\end{equation}
where $C(r,n)$ and $D(r,n)$ are constants. We have tabulated some of these coefficients, along with the values of entropy in the Table \ref{CDTable}. The generic values of $C(r,n)$ and $D(r,n)$ for $n \geq 3$ can be calculated using the equation \eqref{CandD}.
\begin{table}[ht]
\centering
\begin{tabular}{|c|c|c|c|c|}
\hline \rowcolor{Gray}
$n$ & $r$ & $C(r,n)$ & $D(r,n)$ & $\text{EE}_{\text{LP}}^{\infty}$ \\
\hline

\multirow{1}{*}{1}
  & 0 & $2\pi^2$ & $2\pi^2$ & $\ln 2$ \\
\hline

\multirow{2}{*}{2}
  & 0 & 0 & 4 & 0\\
  & 1 & 0 & 4 & 0\\
\hline

\multirow{3}{*}{3}
  & 0 & $3/2$ & $3/2$ & $\ln 2$ \\
  & 1 & $3/2$ & $3/2$ & $\ln 2$ \\
  & 2 & $3/2$ & $3/2$ & $\ln 2$ \\
\hline

\multirow{4}{*}{4}
  & 0 & 2 & 2 & $\ln 2$ \\
  & 1 & 2 & 0 & 0 \\
  & 2 & 2 & 2 & $\ln 2$ \\
  & 3 & 2 & 0 & 0 \\
\hline

\multirow{5}{*}{5}
  & 0 & $(5+\sqrt{5})/4$ & $(5+\sqrt{5})/4$ & $\ln 2$ \\
  & 1 & $(5+\sqrt{5})/4$ & $(5+\sqrt{5})/4$ & $\ln 2$ \\
  & 2 & $(5-\sqrt{5})/4$ & $(5-\sqrt{5})/4$ & $\ln 2$ \\
  & 3 & $5/2$ & $5/2$ & $\ln 2$ \\
  & 4 & $(5-\sqrt{5})/4$ & $(5-\sqrt{5})/4$ & $\ln 2$ \\
\hline

\multirow{6}{*}{6}
  & 0 & 3 & 1 & $\ln[4/3^{3/4}]$ \\
  & 1 & 0 & 3 & 0 \\
  & 2 & 3 & 1 & $\ln[4/3^{3/4}]$ \\
  & 3 & 0 & 1 & 0 \\
  & 4 & 3 & 3 & $\ln 2$ \\
  & 5 & 0 & 1 & 0 \\
\hline
\end{tabular}
\caption{Numerical values of $C(r,n)$ and $D(r,n)$ and large $k$ limit of entropy for low values of $n$.}
\label{CDTable}
\end{table}

At the end, we would like to point out certain identities involving $C(r,n)$ and $D(r,n)$. Here they are.
\begin{itemize}
    \item For $n=\text{odd}$, we will have: $C(r,n)=D(r,n)$ for any $r$.  \\
\textbf{Proof:} Let $n$ be an odd integer. So, $\left\lfloor (n-1)/2\right\rfloor = \left\lfloor n/2\right\rfloor =  (n-1)/2$. Now consider the coefficient $D(r,n)$ and let us change the summation variable to $s'= \frac{n+1}{2} -s$. So we will have:
\begin{equation}
D(r,n) = \dfrac{8}{n} \abs{\sum\limits_{s'=1}^{\frac{n-1}{2}}  e^{-\frac{i \pi  (r+2) (2s'-n)^2}{2n}} \sin ^2\left(\frac{2\pi s'}{n}-\pi\right)}^2  = \dfrac{8}{n} \abs{\sum\limits_{s'=1}^{\frac{n-1}{2}}  e^{-\frac{2i \pi  (r+2) s'^2}{n}} \sin ^2\left(\frac{2\pi s'}{n}\right)}^2 ~
\end{equation} 
where we expanded $(2s'-n)^2=4s'^2+n^2-4ns'$ and notice that only the first term will appear in the phase. This final expression is identical to $C(r,n)$, and hence the proof.
\item For $n=\text{even}$, one of the $C(r,n)$ or $D(r,n)$ will be 0 for odd values of $r$.\\
\textbf{Proof:} Let $r$ be an odd integer where $r=0,1,2,\ldots,n-1$. Also, let $n=2u$. We can write the coefficients as $C(r,n)=\frac{8}{n}\abs{S_C}^2$ and $D(r,n)=\frac{8}{n}\abs{S_D}^2$ where:
\begin{equation}
S_C = \sum\limits_{s=1}^{u-1}  e^{-\frac{ i \pi  (r+2) s^2}{u}} \sin ^2\left(\frac{\pi  s}{u}\right) \quad;\quad S_D = \sum\limits_{s=1}^{u}  e^{-\frac{i \pi  (r+2) (1-2 s)^2}{4u}} \sin ^2\left(\frac{\pi -2 \pi  s}{2u}\right) ~.                                                                                                                         
\end{equation} 
In the sum $S_C$, change the variable to $s'=u-s$. Further, in $S_D$, change the variable to $s'=u+1-s$. In new variables, we can simplify the phases and the $\sin$ terms. Comparing it with the $S_C$ and $S_D$ written in old variable $s$, we obtain the following identities:
\begin{equation}
S_C = e^{-i \pi  r u}\, S_C \quad;\quad S_D = e^{-i \pi  r (u+1)}\, S_D ~.
\end{equation} 
Since $r$ is odd, we clearly see that $S_C$ vanishes when $u$ is odd, and $S_D$ vanishes when $u$ is even. Thus, one of the $C(r,n)$ or $D(r,n)$ will be 0 when $r$ is odd, and $n$ is even. 
\end{itemize}
The above two identities tell us that $\text{EE}_{\text{LP}}^{\infty} = \ln 2$ for odd values of $n$. Further, for $n$ even and $k$ odd values, $\text{EE}_{\text{LP}}^{\infty} = 0$. These were expected from our earlier Result-3.


\section{Conclusion} \label{sec3}
In this work, we have analyzed the large-party limit of the entanglement entropies associated with generic torus link states $\ket{T_{dm,dn}}$. These quantum states are prepared in three-dimensional Chern-Simons theory with compact gauge group $G$ at level $k$. The state $\ket{T_{dm,dn}}$ is associated with the link complement $S^3 \backslash T_{dm,dn}$ and can be computed explicitly using group-theoretic and topological methods. These are $d$-party quantum states living in a $d$-fold tensor product of finite-dimensional Hilbert spaces. 

The entanglement measures associated with this state do not depend on the choice of bipartition. For simplicity, we therefore consider the bipartition $(1|d-1)$, where we trace out $(d-1)$ parties, leaving a reduced system consisting of a single party. We compute the entanglement entropy for this reduced system and analyze its behavior as the total number of parties tends to infinity, i.e., in the limit $d \to \infty$, which we refer to as the large-party limit.

We have shown that, in this regime, the contributions to the entanglement measures arise exclusively from Abelian anyons, while the non-Abelian sectors are suppressed. This leads to a universal upper bound on the entanglement entropy in the large-party limit, given by $\ln |Z_G|$, where $Z_G$ denotes the center of the gauge group.

We illustrated these general results explicitly for the case $G=\mathrm{SU}(2)$, where the large-party behavior of the entanglement entropy for the state $\ket{T_{d,dn}}$ was obtained. After taking the large-party limit, we further examined the semiclassical ($k \to \infty$) behavior of the entropies and found that the resulting values remain finite and consistent with the dominance of Abelian sectors.

These results provide new insight into the large-party structure of quantum entanglement in topological quantum field theories and clarify the role played by Abelian sectors. They also suggest an interesting perspective on multipartite entanglement in Chern–Simons theories. It is well known that non-Abelian anyons, characterized by quantum dimensions greater than one, are essential for universal topological quantum computation, as they support nontrivial braiding statistics and internal fusion degrees of freedom, whereas Abelian anyons do not. Our results indicate that in the large-party limit, the entanglement structure is effectively governed by Abelian sectors, with non-Abelian contributions strongly suppressed. This suggests that, in highly multipartite regimes, the accessible entanglement structure becomes effectively simpler, being controlled by sectors with trivial internal degeneracy. From this viewpoint, the large-party limit may place intrinsic constraints on the complexity of topological quantum states that can be realized or probed through such link constructions. It would be interesting to explore whether this suppression has implications for the scalability of topological quantum information processing in related settings.

We would also like to point out another interesting case study \cite{Balasubramanian_2025} in which the topological entanglement receives contributions only from Abelian anyons. In \cite{Balasubramanian_2025}, the authors study quantum states associated with link complements that are topologically described as circle fibrations over a Seifert surface. The Seifert surfaces are homeomorphic to $\Sigma_{g,n}$, a genus-$g$ Riemann surface with $n$ boundaries and Euler characteristic $\chi = 2 - 2g - n$. These fibered link complement states correspond to $n$-party quantum states. It was shown that in the limit $\chi \to -\infty$, the entanglement structure is governed by Abelian anyons, while non-Abelian sectors are suppressed. This limit can be achieved either by taking $n \to \infty$ (large-party limit) or $g \to \infty$ (large-genus limit). It would be interesting to explore this connection further, as it suggests that limits corresponding to repeated topological composition—either through increasing the number of components or the genus—tend to suppress non-Abelian sectors, which typically carry larger internal degeneracies.

\vspace{0.5in}

\noindent \textbf{Acknowledgment}\\
SD is supported by the “SERB Start-Up Research Grant SRG/2023/001023”.

\newpage
\appendix

\section{Poisson summation formula}
\label{appA}
Let $h(y)$ be a complex-valued smooth function in the domain $y \in \mathbb{R}$. Then, the Poisson summation formula tells us that:
\begin{equation}
\sum_{b=-\infty}^{\infty} h(b) = \sum_{s=-\infty}^{\infty} H(s) ~,
\end{equation}
where $H$ is the Fourier transform of $h$ defined as:
\begin{equation}
H(f) = \int_{-\infty}^\infty h(y)\, e^{-2\pi i f y} dy ~.
\end{equation}
Let us consider the following form of the function:
\begin{equation}
h(y) = f(y/K)\, e^{iK\, \phi(y/K)} ~,
\end{equation}
where $f(y)$ and $\phi(y)$ are two functions and $K$ is a positive constant. Then, the Poisson summation formula becomes:
\begin{equation}
\sum_{b=-\infty}^{\infty} f(b/K)\, e^{iK\, \phi(b/K)} = \sum_{s=-\infty}^{\infty}\, \int_{-\infty}^\infty f(y/K)\, e^{iK\, \phi(y/K)}\, e^{-2\pi i s y} dy  ~.
\end{equation}
Now we are interested in the large $K$ behavior of the following sum:
\begin{equation}
S = \sum_{a=0}^{K-2} f(a/K)\, e^{iK\, \phi(a/K)}  ~.
\end{equation}
So instead of doing a sum over all integers, we want to restrict the sum from $0$ to $K-2$. To achieve our goal, we introduce a smooth cut-off function $\chi(y/K)$ which takes the value 1 on $y \in [0, K-2]$ and dies off outside this region. So the Poisson summation gives:
\begin{equation}
S = \sum_{s=-\infty}^{\infty}\, \int_{-\infty}^\infty f(y/K)\,\chi(y/K)\, e^{iK\, \phi(y/K)}\, e^{-2\pi i s y} dy  ~.
\end{equation}
Let us change the variable $x\equiv y/K$ in the integral and assume large $K$ limit . Now the cut-off function $\chi(x)=1$ for $x\in [0,1]$ and vanishes outside. This will give us the desired formula:
\begin{equation}
\sum_{a=0}^{K-2} f(a/K)\, e^{iK\, \phi(a/K)} = K \sum_{s = -\infty}^\infty\, \int_{0}^1 f(x)\, e^{iK\, \phi_s(x)} dx  ~.
\label{LargeKformula}
\end{equation}
where $\phi_s(x) = \phi(x) - 2\pi s x$. The next step would be to analyze the large $K$ behavior of the integrals on the RHS, which we study as follows.
\section{Asymptotic expansion using the method of stationary phase}
\label{appB} 
Consider the integral:
\begin{equation}
I = \int_{a}^b f(x)\, e^{iK\, \phi(x)}\, dx ~,
\end{equation}
where $f(x)$ and $\phi(x)$ are smooth functions and $\phi'(x)$ has only finite number of zeros in the interval $[a,b]$. In other words, the function $\phi(x)$ has finitely many stationary points in $[a,b]$. The method of the stationary phase \cite{Copson_1965} tells us that the leading order contribution in the large $K$ expansion of $I$ will come from the immediate neighborhoods of stationary points in $[0,1]$ of the phase function $\phi(x)$. Let us analyze the asymptotics for a given stationary point $x_0 \in [a,b]$. There can be three possibilities: $x_0=a$ or $a<x_0<b$ or $x_0=b$. We will analyze each of these possibilities one by one.
\subsection*{Case-1: When $x_0=a$ is the point of stationary phase}
In the neighborhood of $x_0=a$, we will have the following Taylor expansions of $f(x)$ and $\phi(x)$:
\begin{equation}
f(x) = \sum_{j=0}^{\infty}\frac{f^{(j)}(a)}{j!}(x-a)^j \quad;\quad \phi(x)=\phi(a)+ \frac{\phi''(a)}{2!}(x-a)^2 + \ldots  ~,
\end{equation}
where we have ignored the higher-order derivative terms in the Taylor expansion of $\phi(x)$. This is because the leading order contribution will only come from the quadratic term in $\phi(x)$. We further assume that $\phi''(a) > 0$ because this will be the case in our calculation. Thus, we can approximate the integral about a small neighborhood around $a$ as:
\begin{equation}
I \approx \int_{a}^{a+\epsilon} \sum_{j=0}^{\infty}\frac{f^{(j)}(a)}{j!}(x-a)^j\, e^{iK \phi(a)} \,e^{iK \frac{\phi''(a)}{2}(x-a)^2} \, dx ~.
\end{equation}
Let us define $y=\sqrt{K \phi''(a)/2}\, (x-a)$ and for large $K$ limit, we will have:
\begin{equation}
I \approx \frac{\sqrt{2}\,e^{iK \phi(a)}}{\sqrt{K\,\phi''(a)}} \sum_{j=0}^{\infty}\,\frac{2^{j/2}\,f^{(j)}(a)}{j!\,(K \phi''(a))^{j/2}} \int_{0}^{\infty} y^j\, e^{i y^2} \, dy ~.
\end{equation}
The integral on the RHS is a generalization of the standard Fresnel integrals, and its values are known and come out to be:
\begin{equation}
\int_{0}^{\infty} y^j\, e^{i y^2} \, dy = \frac{1}{2}\, \Gamma\left(\frac{j+1}{2}\right)\, e^{i \pi (j+1)/4} \,  ~.
\end{equation}
Substituting the values, we get the large $K$ expansion as:
\begin{equation}
I \sim \frac{\sqrt{2}\,e^{iK \phi(a)}}{\sqrt{\phi''(a) }}  \left[ \frac{\sqrt{\pi }\, e^{\frac{i \pi }{4}} f(a)}{2\, K^{1/2}} + \frac{i\sqrt{2}\, f'(a)}{2 \sqrt{\phi''(a)}\,K} + \frac{\sqrt{\pi }\, e^{\frac{3i \pi }{4}} f''(a)}{4\,\phi''(a)\,K^{3/2}} + \cdots  \right] ~.
\label{x0LeftSP}
\end{equation}
\subsection*{Case-2: When $x_0$ is the point of stationary phase and $a< x_0 < b$}
In the neighborhood of $x_0$, we will have the following Taylor expansions of $f(x)$ and $\phi(x)$:
\begin{equation}
f(x) = \sum_{j=0}^{\infty}\frac{f^{(j)}(x_0)}{j!}(x-x_0)^j \quad;\quad \phi(x)=\phi(x_0)+ \frac{\phi''(x_0)}{2!}(x-x_0)^2 + \ldots  ~,
\end{equation}
where once again, we have ignored the higher-order derivative terms in the Taylor expansion of $\phi(x)$. We further assume that $\phi''(x_0) > 0$. Thus, we can approximate the integral about a small neighborhood around $x_0$ as:
\begin{equation}
I \approx \int_{x_0-\epsilon}^{x_0+\epsilon} \sum_{j=0}^{\infty}\frac{f^{(j)}(x_0)}{j!}(x-x_0)^j\, e^{iK \phi(x_0)} \,e^{iK \frac{\phi''(x_0)}{2}(x-x_0)^2} \, dx ~.
\end{equation}
Let us define $y=\sqrt{K \phi''(x_0)/2}\, (x-x_0)$ and for large $K$ limit, we will have:
\begin{equation}
I \approx \frac{\sqrt{2}\,e^{iK \phi(x_0)}}{\sqrt{K\,\phi''(x_0)}} \sum_{j=0}^{\infty}\,\frac{2^{j/2}\,f^{(j)}(x_0)}{j!\,(K \phi''(x_0))^{j/2}} \int_{-\infty}^{\infty} y^j\, e^{i y^2} \, dy ~.
\end{equation}
The integral on the RHS is again a generalization of the standard Fresnel integrals, and we obtain:
\begin{equation}
\int_{-\infty}^{\infty} y^j\, e^{i y^2} \, dy =  \begin{cases}
    0 & \text{,\, for $j=$ odd} \\ \Gamma\left(\frac{j+1}{2}\right)\, e^{i \pi (j+1)/4} & \text{,\, for $j=$ even}
\end{cases} ~.
\end{equation}
Substituting the values, we get the large $K$ expansion as:
\begin{equation}
I \sim \frac{\sqrt{2}\,e^{iK \phi(x_0)}}{\sqrt{\phi''(x_0) }}  \left[ \frac{\sqrt{\pi }\, e^{\frac{i \pi }{4}} f(x_0)}{K^{1/2}} + \frac{\sqrt{\pi }\, e^{\frac{3i \pi }{4}} f''(x_0)}{2\,\phi''(a)\, K^{3/2}} + \cdots  \right] ~.
\label{x0SP}
\end{equation}
\subsection*{Case-3: When $x_0=b$ is the point of stationary phase}
In the neighborhood of $b$, we will have the following Taylor expansions of $f(x)$ and $\phi(x)$:
\begin{equation}
f(x) = \sum_{j=0}^{\infty}\frac{f^{(j)}(b)}{j!}(x-b)^j \quad;\quad \phi(x)=\phi(b)+ \frac{\phi''(b)}{2!}(x-b)^2 + \ldots  ~,
\end{equation}
where we have ignored the higher-order derivative terms in the Taylor expansion of $\phi(x)$, and we will further assume that $\phi''(b) > 0$. Thus, we can approximate the integral about a small neighborhood around $b$ as:
\begin{equation}
I \approx \int_{b-\epsilon}^{b} \sum_{j=0}^{\infty}\frac{f^{(j)}(b)}{j!}(x-b)^j\, e^{iK \phi(b)} \,e^{iK \frac{\phi''(b)}{2}(x-b)^2} \, dx ~.
\end{equation}
Let us define $y=\sqrt{K \phi''(b)/2}\, (x-b)$ and for large $K$ limit, we will have:
\begin{equation}
I \approx \frac{\sqrt{2}\,e^{iK \phi(b)}}{\sqrt{K\,\phi''(b)}} \sum_{j=0}^{\infty}\,\frac{2^{j/2}\,f^{(j)}(b)}{j!\,(K \phi''(b))^{j/2}} \int_{-\infty}^{0} y^j\, e^{i y^2} \, dy ~.
\end{equation}
The integral on the RHS can be obtained as:
\begin{equation}
\int_{-\infty}^{0} y^j\, e^{i y^2} \, dy =  \frac{1}{2}\Gamma\left(\frac{j+1}{2}\right)\, e^{i \pi (5j+1)/4} ~.
\end{equation}
Substituting the values, we get the large $K$ expansion as:
\begin{equation}
I \sim \frac{\sqrt{2}\,e^{iK \phi(b)}}{\sqrt{\phi''(b) }}  \left[ \frac{\sqrt{\pi }\, e^{\frac{i \pi }{4}} f(b)}{2\, K^{1/2}} - \frac{i\sqrt{2}\, f'(b)}{2 \sqrt{\phi''(b)}\,K} + \frac{\sqrt{\pi }\, e^{\frac{3i \pi }{4}} f''(b)}{4\,\phi''(b)\,K^{3/2}} + \cdots  \right] ~.
\label{x0RightSP}
\end{equation}

\bibliographystyle{JHEP}
\bibliography{Arxiv}

\end{document}